\documentclass[english,12pt]{article}

\usepackage{amssymb}
\usepackage{graphicx}
\usepackage{amsfonts}
\usepackage{bbm}
\usepackage[T1]{fontenc}
\usepackage[latin1]{inputenc}
\usepackage{babel}
\usepackage{color}

\newcommand{\be}{\begin{equation}}
\newcommand{\ee}{\end{equation}}
\newcommand{\rf}[1]{(\ref{eq:#1})}
\newcommand{\X}{\bar{\mathcal{X}}}
\newcommand{\Z}{\mathcal{Z}}
\newcommand{\R}{\mathcal{R}}
\newcommand{\G}{\mathcal{G}}
\newcommand{\M}{\mathcal{M}}

\begin{document}

\title{Massive Kaluza-Klein Gravity}

\author{D. C. N.  Cunha \thanks{disrael@fis.unb.br}
\hspace{2mm},   M. D. Maia\thanks{maia@unb.br} \\ Universidade de Brasilia,\\ Instituto de Física, 70910-900 Brasilia  D.F.
}

\maketitle

\begin{abstract}

The  non-Abelian Kaluza-Klein unification  of  gravitation  with gauge  fields  theory is   reformulated,  with the inclusion of a  massive spin-2 field  defined by the  extrinsic  curvature. The  internal  space is non-compact,  characterized  by the  group of  rotations of  vectors  orthogonal to the space-time.
The  non-compactness of  the  internal  space warrants  the  solution of  the    fermion  chirality  problem  of  the original  Kaluza-Klein  theory  and makes it   closer  to  the  more  recent  Brane World  paradigm, in  special  to the so called  DGP     model. However,  the access of  gravitation to  the extra dimensions  is   defined by  the  mentioned    massive spin-2  field obeying the  Fierz-Pauli   equation.   The  existence of  a short range  gravitational  component makes possible  to  apply  the modified  Kaluza-Klein  unification   to  the Tev  scale of  energies.

\end{abstract}

\section{Introduction}
The  recent  detection of  the Higgs  particle  at the LHC gives  a  new  support  to  the  standard model   of  the  fundamental interactions.  It also hints   that    gravitation as  the  force  acting on masses  should  be   somehow  included  in  that   model.  One  strong  proponent of such  unification  is  the   Kaluza-Klein  theory,   in  which   gauge  fields  are  part of  a  higher dimensional  metric space  obeying the Einstein-Hilbert principle.

Gravitation  has been traditionally  neglected  in the  standard model  because    the  so  called   gravitational  hierarchy,  whereby  the  gravitational field  is  too  weak to play a significant  role  in presence  of   gauge  fields,  except  perhaps   at  the  Planck  regime. This   hierarchy is  a  consequence  of   the presence of  Newton's  gravitational  constant  $G$   in  Einstein's  equations. However, in   a    higher-dimensional  gravitational theory  the   coupling constant  of  gravitation  with  gauge  fields  and  matter sources  cannot be  defined by Newton's  constant,  simply because  that  constant depends on the dimensions  of the  space,  namely  3  space  dimensions. Therefore, in  higher  dimensions   the gravitational hierarchy  may be  broken.

Independently  of  how  such  higher  dimensional theory is  formulated,
it  must be  compatible  with   General Relativity  which provides  the  link  with   Newton's gravity.  Therefore,   we  face   the possibility  that  gravitation may  be more  complex  than  the  current  understanding  of  such  force, as for example by  providing  two  levels  of   interaction: one  being given  by  the traditional   massless   Einstein-like  gravitation  responsible  for   the  classical long  range  interactions,  which  we  experience in our everyday  life.  The other    is a  massive   strong gravity  acting at   short  range,   at the Tev   scale of  energies. Of the  later  we  know  very little, except perhaps   a  few  hints  from  the  new  experimental high energy physics at the Tev scale. It is  also possible  that  it has  implications  to    extragalactic   astrophysics.

Massive spin-2  fields   were     described by  the  Pauli-Fierz  action   in   1939  and  its  interaction with   four-dimensional Einstein's   gravitation  were  considered  by  various  authors  in the  70's,     including  the  possibility of   existence of   a  short range   strong   gravitational    component   and  its  effects  on  high  energy  hadron physics,  but also as  a possible  explanation  of  the   acceleration of the universe, avoiding   the problems  associated  with  the cosmological  constant \cite{FierzPauli, Gupta,Salam}.  The purpose of  this  note  is  to   revise  Kaluza-Klein theory  with  the two   new  ingredients. Firstly,  the internal space is  not  compact.  Secondly   the propagation of  gravitation  in   the extra-dimensional space  is  a  consequence  of  a   massive  component of  gravitation.  Let us  start   with  a  very brief  review  of  the  original non-Abelian  Kaluza-Klein theory.

During  the   subsequent  20 year period after it  was  firstly proposed in 1963,   the non-Abelian Kaluza-Klein theory   was considered   to be a   serious candidate  for  a  successful unification  of the  four fundamental  interactions (gravitation and the  standard  gauge  forces)  at   the Planck scale. The  theory  was    defined in   a  higher-dimensional space  with  product  topology  $V_4\times  B_N$,  where  $V_4$  is  the space-time  and  $B_N $  is  a  compact  internal space with  typical diameter  of  the  order of Planck's length. The   metric geometry of  the   total  space is  defined by  the   Einstein-Hilbert  action,   that   could  be   decomposed in the    four-dimensional  Einstein-Hilbert  action defined  in the space-time $V_4$,  plus  the  Yang-Mills  action  defined  with respect  to  the  group  of  symmetries of $B_N$.

 Then, in  1984 a  contradiction  of the  theory was noted,  whereby its  predictions at the electroweak limit of  Tev   scale  of energies  did  not  agree with  the observed  chiral motion of  fermions.
The incompatibility  with the   fermion chirality  was soon    understood  to be  a     consequence of the  mathematical structure of the  higher  dimensional space,  which was postulated to be the topological product $V_4\times  B_n$  and  the  small  size of  $B_n$.  This    assumption was  justified  by  the necessity  that the  extra  dimensions  could not be  directly observed  at any  practial   level of  energy.  It was  understood  that  $B_n$  also  introduced  an  additional mass to fermions in the theory,  proportional  to the inverse of  Planck's length.

In  spite of the many efforts to  save  the  theory  presented  at that  time  \cite{Wetterich,Weinberg,Maia,Wesson},  it was practically   abandoned.  The   general  feeling  was that,  even  if  the  fermion chirality problem  could  be  solved,  we  would  still  facing  an  even  more  difficult  problem   which was  the  quantization of  a   theory  based on the Einstein-Hilbert  principle.

However, the legacy of  the Kaluza-Klein theory  to  theoretical physics  is considerable,  mainly because it opened a perspective for the existence  of  extra  dimensions, beyond  the  four  dimensions  of  the   space-time. The   literature produced on  Kaluza-Klein theory is  vast,  and  from   these  we  have  much  to learn.  For historical and technical  reviews  we  suggest a  look at \cite{Witten,Chodos,Mecklemburg}. A  substantial  contingent  of researchers   moved  to the then  infant  theory of  strings also based  on the property of  submanifolds  embedded  in  a  space with more than  four  dimensions.

The   quantization of  the geometry  in  Kaluza-Klein  theory  would be  unavoidable  because  it  incorporates  a  quantizable  component, in  the  form of  a   gauge  field.  The  presence of  such  field  built in the metric  geometry   of  the  total space  implies  that the remaining   components  of the metric   also  would show  some   quantum  fluctuations.  Here  we  are using the  same principle   proposed by  Ashtekar  to  use  the group  of  holonomy  of  the   triads  to  get  an auxiliar $SU(2)$ field  which  would  induce  quantum  fluctuations in the  remaining  geometry  (This  later     developed  into the presently  very active loop quantum gravity program) \cite{Ashtekar}.  The difference is  that  here  the gauge  fields  are  already  present  as  additional  components of the Kaluza-Klein metric,  so that   we do not  depend  on the  holonomy  groups  and  its  implications to the  use of  Wilson's  integral.

However, if  a  unification theory like   Kaluza-Klein theory  is  to be effective,  the  hierarchy  of  the  gravitational  field  must be   resolved.       Newton's  gravitational  constant $G$  was  derived  from observations within  the  context  of  Newtonian  mechanics,   measured  with   absolute  time  separated  from the  three-dimensional  distances. Therefore,   $G$   is  consistent  with the   topology of  the Newtonian  space-time  given by the  product $I\!\! R^3 \times I\!\! R$,  so that
 the     physical  dimensions  of   $G$     are  appropriate to  convert  the squared mass  by   the squared  three-dimensional distance   into  the Newtonian gravitational  force.
 When   the  same  constant is   imported into  General  Relativity  its physical  dimensionality   does not change and  it remains  entirely  compatible  with the  fact that    Einstein's  equations  are of  hyperbolic nature,   implying that  the    topology  of  the  space-times is  $I\!\! R^3 \times I\!\! R$   \cite{Bruhat}.  This is  the same  topology of the Newtonian  space-time, the  difference  being that  in General relativity  the  time is  local  and  the space sections  are not simultaneous  sections.   In spite of  this,   the  value  of  $G$ determined  by  Newtonian  mechanics,  imposes   an  enormous energy   difference  between the  energy levels  of  Einstein's  gravitation  and those of  the   relativistic  gauge fields.

On  the other hand,  as  a consequence  of  the  mathematical  structure of Maxwell's  equations,   or  more generally   of the  Yang-Mills  equations,  together  with  the  experimental evidences  of  the  standard  model of particles  and  fields,   the topology of  the (Minkowski) relativistic  space-time is not   separated  as  a  product,  but  has  an  inseparable  relation  between  time  and  space  in a  four dimensional integrated  space-time.

Indeed, the gauge  field   strength $F=F_{\mu\nu}dx^\mu \wedge dx^\nu$ is    a 2-form (or equivalently, a covariant  rank 2 antisymmetric Maxwell-like tensor).  Its  components  are $F_{\mu\nu}=[D_\mu , D_\nu]$,  where   $ D_\mu =\partial_\mu + A_\mu$  is  the  gauge  covariant derivative with respect to the  components  of the  gauge potential   $A_\mu$,   written in the   adjoint  representation of  the local gauge  symmetry \cite{Huang,Moriyasu,MaiaGFI}.  Using the notation of  exterior product   the  Yang-Mills   equations is  written as
$$D \wedge F^*=0\;\;\; \mbox{and}\;\;\;    D\wedge F  =4\pi j$$
where the  star   denotes  the   dual
$F^*   =F^{*}_{\mu\nu} dx^\mu \wedge dx^\nu,\;\; F^*_{\mu\nu} =\epsilon_{\mu\nu\rho\sigma}F^{\rho\sigma}$.
Therefore,   the  3-form  $D\wedge F $  must  equal  to the    current one-form. However,   three-forms  and  one-forms are  isomorphic   only in  four  dimensional spaces.
On the other hand,   the gravitational  field  defined by Einstein's  equations  does not  have the  same  type of  gauge  confinement  to the  four-dimensional  space-time based on the  dynamical  equations. Therefore,  differently from gauge  fields, in principle  the gravitational  field   may  propagate  also  along  extra dimensions  if they   exist.

Under  the  hypothesis of   the  space-time  being a  subspace  of  a larger host    space  defined by  the   Einstein-Hilbert principle,  the  host  space   have  the  same  hyperbolic  characteristics, but  where  with   the associated  topology  is  decomposed in the product  $V_4 \times  \M_N$,    where    $V_4$  is  a  four-dimensional physical space-time  and  $\M_N$  is the  N-dimensional local orthogonal  space  generated by  the  extra  dimensions.

\section{Smoothing the  Space-time}

The  characterization of  Riemannian  manifolds as topological  spaces  was established  by  O. Veblen and  H.  Whitehead,  77  years  after Riemann's  paper and  25  years  after  Einstein's  use of  the Riemannian geometry to describe gravitation.  Such  time gap reflects  a  degree   of  conceptual complexity   in  Riemann's  original paper \cite{Riemann,Nachlass,Scholz}.  Another  topological  characteristic  in Riemann's  paper which  remained  obscure for  some  time was  the notion of the  shape   of  the  manifold: In  his  presentation,  Riemann  commented   that his  geometry  was not  capable  to   distinguish  between  two  different  manifolds  with  zero  curvature,  as  for  example  between  a  plane and  a cylinder among an  infinite  choice of non-trivial  flat  Riemann manifolds. This  topological  deficiency  cannot be  ignored  in Einstein's  gravitational theory  because the  flat  space-time acts also as  a  ground  state  for  gravitation.

A  solution  of  such shape  problem   was conjectured  in  1871  by  L. Schl\"afli,  suggesting that  the notion  of  shape of  an observed object  cannot  be decided intrinsically.  He proposed that  the Riemann curvature  could provide an  unambiguous  measure of  shape  provided it  could be     compared  with  the curvature  of  another  Riemannian manifold.  In principle  any  other Riemannian   manifold  could act  as a reference of  shape as long as  both  manifolds  could be somehow locally compared.   This  would  require  that,   like the old  Euclidean geometry,  any  n-dimensional  Riemannian  manifold could  be  locally  embedded in another  D-dimensional  manifold.  The  embedding manifold itself could  act as    a  background  reference  for  shape.  Although  Schl\"afli's solution is   very intuitive,    it  took  a  while until a complete  formulation of  the problem,  which  finally  culminated  in the  derivation of  the Gauss-Codazzi-Ricci  equations for  the  embedding.
These  equations  involve   not only the metric  but also  the other two  fundamental  forms  of  differential geometry.

In  General relativity  the   definition of  shape is  local,  so that  the mentioned  embedding is also only local,  characterized by  a   map  which takes a neighborhood  of  a point of  manifold $\bar{V}_n$  into  the  embedding space  $V_D$.  In the case of  a  space-time  the local  embedding  is
\[
\X:  \bar{V}_4 \rightarrow  V_D
\]
such  that its    components  $\X^A,\;\; A=1 \cdots D$  are  functions  of  the  space-time  coordinates.  These functions  must be    differentiable  and  regular,   so that $\bar{\X} $  can be   locally  invertible, thus   enabling the  local   recovery  of the  original space-time. Since  the line  element in  space-time   is  the  same, independently  of the fact that  the  space-time  is  embedded or  not,  we must have
\be
ds^2 =
 {\cal G}_{AB}\X^A_{,\mu}\X^B_{,\nu}dx^\mu dx^\nu=\bar{g}_{\mu\nu}dx^\mu dx^\nu  \label{eq:ds}
\ee
where   ${\cal G}_{AB}$  is  the metric  of  the embedding space  and  $\bar{g}_{\mu\nu}$ is  the metric of  the   space-time.
The  derivatives $\X^A_{,\mu}$   define  a  basis of  the tangent space to the four-dimensional  embedded  space-time.     To  complete  the  basis  of  the  embedding space we  need  an  additional    $N=D-4$   vectors $\eta^A_a$ which can be  chosen  to be orthogonal  to  $V_4$ and to themselves  at  each point.  In  this  way  we obtain a Gaussian  reference frame  in the embedding  space     $\{\X^A_{,\mu}, \bar{\eta}^A_{a}\}$ such that
(Hereafter,  Greek  indices $\mu ,\nu ...$  run  from  1  to  4, capital case  Latin indices  run  from 1 to D and  small case Latin indices  run from  5  to D):
\be
{\X}^{A}_{,\mu}{\X}^{B}_{,\nu}{\cal G}_{AB} =\bar{g}_{\mu\nu},\;\; {\X}^{A}_{,\mu}\bar{\eta}^{B}_{a}{\cal G}_{AB}=0,\;\;
\bar{\eta}^{A}_{a}\bar{\eta}^{B}_{b}{\cal G}_{AB}=\bar{g}_{ab} \label{eq:X}
\ee
where  $\bar{g}_{ab}=\epsilon\delta_{ab}$,  where  $\epsilon =\pm 1$ which defines  the  signature of  the extra  dimensions.

Writing  the Riemann  tensor  of  $V_D$   in   this  Gaussian frame  and   applying  all  index symmetries  and    curvature identities,   the only remaining independent equations  are  the    well known Gauss-Codazzi-Ricci equations. These form  the  integrability conditions  for  the  embedding,  but essentially they are  the  components  of the Riemann tensor  written in the Gaussian frame \cite{Eisenhart}:
\begin{eqnarray}
&&\R_{ABCD}{\X}^{A}_{,\alpha}
{\X}^{B}_{,\beta}{\X}^{C}_{,\gamma}{\X}^{D}_{,\delta}
=\bar{R}_{\alpha\beta\gamma\delta} - 2\bar{g}^{mn}\bar{k}_{\alpha[\gamma
m}\bar{k}_{\delta]\beta n}
 \label{eq:Gauss}\\
&&\R_{ABCD} {\X}^{A}_{,\alpha}
\eta^{B}_{b}{\X}^{C}_{,\gamma}{\X}^{D}_{,\delta}
=\bar{k}_{\alpha[\gamma b; \delta]} -
\bar{g}^{mn}\bar{A}_{[\gamma mb}\bar{k}_{\alpha\delta]n }\label{eq:Codazzi}\\
&&\R_{ABCD}\eta^{A}_{a}\eta^{B}_{b} {\X}^{C}_{,\gamma} {\X}^{D}_{,\delta} = -2\bar{g}^{mn}A_{[\gamma
ma}\bar{A}_{\delta]n b} -2\bar{A}_{[\gamma a b ; \delta]} - \bar{g}^{mn}\bar{k}_{[\gamma m
a}\bar{k}_{\delta]nb} \label{eq:Ricci}
\end{eqnarray}
where  $\bar{k}_{\mu\nu a}$ denote the components of the  extrinsic curvature, one for each  extra dimension  with label ``a'',    and $\bar{A}_{\mu a b} $ denote the components of the third fundamental form defined   respectively  by
\be
\bar{k}_{\mu\nu a} =- {\cal G}_{AB}\eta^A_{a,\mu}\X^B_{,\nu},\;\;\;\;  \bar{A}_{\mu a b}=  {\cal G}_{AB} \eta^A_{a,\mu}\eta^B_{b}   \label{eq:kA}
\ee
Notice  that  equations (\ref{eq:Gauss}-\ref{eq:Ricci})  represent only the  conditions  for the  existence  of  the  embedding of  a  space-time.  No   boundary  conditions  were provided to guarantee  that the  embedding  is  unique\footnote{In the   Randall-Sundrum  brane-world  model,  the  space-time  is  embedded in the  five  dimensional anti deSitter space $AdS_5$. It is assumed  that the space-time  acts  as  a mirror  boundary   for the  higher-dimensional  gravitational  field.  This condition   has  the   effect  that  the extrinsic  curvature becomes an  algebraic function of  the   energy-momentum tensor of matter    confined  to the four-dimensional embedded space-time. In  more than  five  dimensions,  which is  our case in  study, that boundary condition  does not make  sense    because the extrinsic  curvature  acquire an internal index while  the energy-momentum tensor   does not  have  this degree of  freedom. As  we  shall see later, the extrinsic  curvature behaves  as  a dynamical field. Finally,  our main objective here   cannot be  accomplished in the  Randall-Sundrum model  because  $\bar{A}_{\mu ab}$ simply  do  not exists  in the  case of  a  single   extra  dimension}.

The  local and  isometric embedding  of  space-times  can be determined  in three   different  ways: First  by   just try and  error,  finding the  functions  $\X^A$ so that \rf{ds}, \rf{X} and  (\ref{eq:Gauss}-\ref{eq:Ricci}) are  satisfied. In this approach  the  normal  vectors  need  to be  calculated in an  ad-hoc manner  so  that  the  second  and  third differential forms may be  determined.   Several  examples of  such procedure  are  given in \cite{Rosen}.  The so called  Penrose   embedding diagrams are simplified  examples  of   such procedure represented  by   two  dimensional  graph  while   the  remaining  dimensions  are taken to be zero.

The  second  procedure consists in analytically solving  the  integrability equations   (\ref{eq:Gauss}-\ref{eq:Ricci})  based  on the  well known  theorems  of  Janet and  Cartan  \cite{Janet,Cartan}.  Since  those  equations   form  a   non-linear system of  equations on  its  three variables  $\bar{g}_{\mu\nu},\;\;  \bar{k}_{\mu\nu a},  \;\; \bar{A}_{\mu  a b}$, the assumption   that   the  functions   $\X^A (x)$ are  analytic, imply that  all fundamental forms  are also analytic  functions. Some  results  are   obtained  for  a local embedding.  For  example  for  a n-dimensional manifold it is  found  that  at most  $ D=n(n+1)/2$ are  required   for  the embedding  space.  As  far  as  the mathematical analysis  on manifolds  with positive defined metrics are  concerned this  is  fine. However, for pseudo Euclidean manifolds not  all theorems  based on  converging positive  power  series  apply. In  addition,  we  remind  that  in the real world the  analytic assumption is   difficult to attain.

The  third  procedure  uses  a non-trivial   theorem  by  John Nash,  stating that  the solution  of  the  equations  (\ref{eq:Gauss}-\ref{eq:Ricci})  was  obtained  from the  supposition that  an initially   given  embedded  manifold (or  better, a  space-time)    $\bar{V}_4$  can  be  \textit{smoothly   deformed}  along  the  normals producing  a  new   embedded  manifold  $V_4$, with  new extrinsic  curvature satisfying the  condition
\be
k_{\mu\nu a} =-\frac{1}{2}\frac{\partial  g_{\mu\nu}}{\partial y^a}   \label{eq:York}
\ee
where  $g_{\mu\nu}$  and  $y^a$  denote respectively  the metric of  the deformed space-time and the    extra-dimensional  coordinates
 \cite{Nash}.

The  smoothing  condition  is  related  to  the  notion of  geometric   deformation,  which  by  turn is  defined by  a   continuous  displacement  of  points   of  the original manifold,
$\bar{V_4}$  following   a  flow  of  lines which  cross   orthogonally the  original manifold  without losing its   continuity and regularity,  by means  of  the Lie transport  \cite{Pirani},  so  it may be called  the Nash  geometric flow.  The  first  report  of  the condition \rf{York}  appeared  in    1920  in a book by  J. Campbell,  postulated  to    prove  that  any  space-time  could be embedded in a Ricci-flat  five-dimensional  space \cite{Campbell}.  However,   along his proof Campbell implicitly  used  the  analytic expansion \cite{Dahia},  so that  implicitly  he  was using  Cartan's  analytic  embedding.  The same   expression   was  independently  derived   by J. York in  1971  to implement  the  ADM   foliation  of  space-time  by 3-dimension al  surfaces  \cite{York}.
To understand  Nash's    smoothing deformation process,  we may  use an  analogy  with another more recent  smoothing process derived by   Richard Hamilton,  and  subsequently   applied  to the    proof  of  the  Poincar\'e conjecture  \cite{Perelmann,RHamilton}.

In the  derivation of  the  heat  equation  J. Fourier used two  approaches  to measure  the  increase of  temperature  in  a  spherical body embedded in a compact solid    situated  near  source of  heat:  On the one  hand, Fourier  considered  the  specific  capacity   of  the  body material  to  absorb  heat  per  unit of  volume  and  time. On the other hand,  he  considered   the  flux of   heat flow lines per unit  of  area  crossing orthogonally  the  body  surface  and  the  embedded  sphere. Assuming that there  are  no  additional  heat sources  or  sinks inside the body, the  comparison  between  the  two measures of  heat led  to  the Fourier's  parabolic  heat  equation
\be
\nabla^2  u  =\frac{\partial  u}{\partial  t} \label{eq:Fourier}
\ee
Between the  initial surface  and  the end  sphere  we may draw  an  infinite  sequence  of  surfaces  always  orthogonal   to  the flow lines.    This procedure may be  photographed  in a  time  sequence  and  the resulting  film  may   be played back, producing  a   continuous deformation of  the   body's  surface, without  singularities  or  cusps, converging  at the end  to  the sphere.

To translate   the  above reasoning  to   Riemannian   (intrinsic) geometry,  consider  the   Ricci tensor wriiten   as
\[
R_{\mu\nu}=  (log\sqrt{g})_{,\mu\nu} -\Gamma^\rho_{\mu\nu;\rho}  + \Gamma^\sigma_{\mu\rho}\Gamma^\rho_{\nu\sigma}
-\Gamma^\rho_{\mu\nu}(log\sqrt{g})_{,\rho}
\]
Therefore,  in  geodesic  coordinates,  the  Ricci scalar becomes
\be
R=  g^{\mu\nu}R_{\mu\nu} =  \nabla^2 (log\sqrt{g})   \label{eq:aux0}
\ee
In the intended   analogy,  instead  of  varying temperature  with time,   there is  a  surface   change in  an orthogonal  direction $y$. Thus, replacing  the temperature  $u$   in \rf{Fourier} by
$u=log \sqrt{g}$,  we obtain
\[
 g^{\mu\nu}R_{\mu\nu} = \nabla^2 ( log\sqrt{g})=  g^{\mu\nu}\frac{1}{2}\frac{\partial g_{\mu\nu}}{\partial t}  \label{eq:aux1}
\]
Comparing    equations   \rf{aux0} and  \rf{aux1}  and  solving the  resulting tensor  equation  in  $R_{\mu\nu}$,    replacing   $t$  by   an  arbitrary  coordinate  $y$,  we obtain    up  to  the addition of a   traceless  tensor,  the   Ricci flow condition (the minus  sign corresponds to having  the   the  heat flow in the  opposite  direction of  the  shrinking  surface.)
\be
R_{\mu\nu}= -\frac{1}{2}\frac{\partial g_{\mu\nu}}{\partial y}
	\label{eq:RicciFlow}
\ee
It is    a  simple  exercise  to see that  this  condition is  not  compatible  with  Einstein's  equations.

The  Nash   deformation of embedded submanifolds  follows  a   similar    picture, but  with the difference  that   flow lines  are  the  orbits  of  the   Lie  transport of  points  in an initial manifold  along  its orthogonal   directions  as  described by  \rf{Z}. However,
the  physical   meaning   of  Nash's  flow  is  not related  to   Fourier's heat equation as    it  is  entirely  geometrical.
To  derive the  Nash's  smoothing condition,  consider  an  infinitesimal displacement   $y^a$  of the  points of  an initial  manifold.  The  Lie  derivative    of the coordinates  $\X^A$ along  the  orthogonal  direction gives
\begin{eqnarray}
\Z^A (x^\mu,y^a) & =&\X^A(x^\mu)  +y^a  (\pounds_{\eta_a} \X)^A    \label{eq:Z}
\end{eqnarray}
Notice  that  $\pounds_{\bar{\eta}} \bar{\eta}=0$,  so that  these  orthogonal vectors   do not propagate. Consequently,   the  above  coordinates  to define   a new  set of  embedding  equations   similar to  those for \rf{X}:
 \be
{\Z}^{A}_{,\mu}{\Z}^{B}_{,\nu}{\cal G}_{AB} =g_{\mu\nu},\;\; {\Z}^{A}_{,\mu}\bar{\eta}^{B}_{a}{\cal G}_{AB}=g_{\mu a},\;\;\bar{\eta}^A_a \bar{\eta}^B_b {\cal G}_{AB}=g_{ab}\label{eq:Z}
\ee
The  deformed  manifold  $V_4$  defined by these  equations has  a new  set of  orthogonal vectors  $\eta_a (x^\mu,y^a)$  which are not   necessarily parallel    to $\bar{\eta}$. Consequently  the new  cross  metric  components  $g_{\mu a}$ defined  above  are  not necessarily zero.

By  repeating  this procedure starting from  $\Z^A_{,\mu}$  and $\eta^A_a$ we  obtain a continuous  sequence of   four-dimensional  embedded  manifolds $V_4$,  each one  being a small deformation of  the preceding one.  From \rf{Z}  we  obtain the  new  fundamental forms
\begin{eqnarray}
{g}_{\mu\nu}(x,y) & = &
{\mathcal{Z}}_{,\mu}^{A}{\mathcal{Z}}_{,\nu}^{B}{\mathcal{G}}_{\!
AB}=\bar{g}_{\mu\nu}\!\!- 2 y^{a}\bar{k}_{\mu\nu a}
+y^{a} y^{b}[\bar{g}^{\alpha\beta}\bar{k}_{\mu\alpha
a}\bar{k}_{\nu\beta b}+g^{cd}\bar{A}_{\mu
ca}\bar{A}_{\nu db}]\vspace{1mm} \label{eq:gmunu}\\
{g}_{\mu a}(x,y) & = &
{\mathcal{Z}}_{,\mu}^{A}{\eta}_{a}^{B}{\mathcal{G}}_{\! AB}=\!\!  y^{b}\bar{A}_{\mu ab}=A_{\mu a},  \label{eq:gmua}\\
{g}_{ab}(x,y) & = &
{\eta}_{a}^{A}{\eta}_{b}^{B}{\mathcal{G}}_{\! AB}= \bar{g}_{ab} \label{eq:gab}\\
{k}_{\mu\nu a}(x,y) & = &
-\eta_{a,\mu}^{A}{\mathcal{Z}}_{,\nu}^{B}{\mathcal{G}}_{AB}
=\bar{k}_{\mu\nu
a}\!\!- y^{b}[\bar{g}^{\alpha\beta}\bar{k}_{\mu\alpha
a}\bar{k}_{\nu\beta
b}-\!\!{g}^{cd}\bar{A}_{\mu ca}\bar{A}_{\nu db}],\hspace{3mm}\label{eq:kmunua} \vspace{1mm}\\
{A}_{\mu ab}(x,y) & = &
\eta_{a,\mu}^{A}\eta_{b}^{B}{\mathcal{G}}_{AB}\!\!=\!\!\bar{A}_{\mu
ab}(x)\label{eq:Amuab}
\end{eqnarray}
Taking the  derivative  of  $g_{\mu\nu}(x,y)$   with  respect  to  $y_a$ in  \rf{gmunu} and  comparing  with    \rf{kmunua},    we obtain  Nash's smoothing  condition  \rf{York}.

The generalization of Nash's  theorem    to  manifolds  with   non-positive  metrics (Lorentzian manifolds) is  well known \cite{Greene,Oneil}.  In  this   generalization, it  is  shown  that the  maximum number  of extra  dimensions  required for the  local  embedding is  $n(n+3)/2$, as opposed to the result  $n(n+1)/2$ predicted  by  the analytic embeddings. Furthermore,  the metric  signature of the embedding space is  not  free  to be  chosen,   but it  depends  on  the   topological  properties of the embedded  space-time.  A  notorious example is  given by the   spherically  symmetric  space-times:  for the Schwarzschild space-time,  the  embedding    signature is  $(4,2)$,   but  for its  geodesically  complete (the maximal  analytical  extension),  known  as   the  Kruskal  space-time, the  signature  changes  to  $(5,1)$  \cite{Fronsdal}. Since  the   Schwarzschild  space-time  is  a  subset of  its geodesically  complete  extension,    this  change  of  signature can  only be  explained   by   the   inclusion  of the  topological   difference between   the  two  space-times: The  Nash  deformations  break  down  at the Schwarzschild  horizon because  the  extrinsic  curvature is  not  defined  there. On the other hand, the Nash embedding  applies  to  the Kruskal metric  from  $r=0$  to  infinity.

As  we  have  already mentioned, the  metric signature  of  the four dimensional  space-times  is  a result  of  the  electromagnetic  theory,   whose  equations are invariant under  the Poincaré  group  in the  four-dimensional Minkowski space-time. Such  perception of  a single-time as represented  by a  negative metric  component  is  carried  over  General  Relativity  without any  further  ado, but  only  to make it  consistent   with the  Minkowski  tangent space postulate of  flat limit.
Because  our gauge  field   probes  are  all  confined, there is   no  experimental   support   which may lead  us  to    conclude  that the  minus signature of  an  extra  dimensions  correspond  to  a  time coordinate.  In  face of  such  doubt, we may  as  in  the theory of  curves,   reparametrize  the  extra  dimensional  orbits   by its  arclength  $y^a\rightarrow s^a$,  in which case  the velocity  vector  has  unit  norm.   If necessary,  it is  possible  to   multiply these  arclengths    by $\frac{1}{c}$  to make  them   truly time-like.

\section{The Kaluza-Klein Geometry}

In  the  60's   D. W. Joseph Y. Ne'emann,   proposed  that the     (internal) gauge groups are  isomorphic to  the  group of  rotations  of the extra  dimensional    space of  an embedding space of the  space-time,   generated  by $N$ vector  fields  orthogonal  to the space-time
\cite{Joseph,Neemann}.   This  is interesting  in the extent  that  all physics  will  be   defined  in the  same   geometric  structure.
In the  application of  this  proposal  to  Kaluza-Klein theory,  those vectors are    not directly observable,  so  that   the justification for a  small  compact internal  space of the original  Kaluza-Klein theory is no longer required. However,  we  will  show  below   that  the  proposed  geometrization of  the   gauge  symmetry  leaves  an observable footprint on the  space time.

  The  only  stable   ground  state for the higher-dimensional  gravitational  field  in the  original  Kaluza-Klein theory, is  the    plane-flat  D-dimensional    space  \cite{Witten}.    Therefore,  it is  reasonable  to  use  the  plane-flat  embedding  space   as  the    ground  state of  the present revision of  Kaluza-Klein theory,  as  the  background reference  for  curvature  and  also   as  the  source of  the internal symmetries as proposed by  Joseph and  Ne'emann.

Since   the metric of the space-time  is  induced  by  the  metric of the   embedding  space,   it foloows  that  the  geometry of  the latter   space  must be  defined  by  the  same  Einstein-Hilbert principle:
\be
\frac{\delta}{\delta {\cal G}_{AB}}\int{({\cal R}-k_{g_*}{\cal L}_m})\sqrt{\cal{G}}d^{D}v =0  \label{eq:EH}
\ee
where  ${\cal R}$  denotes  the higher-dimensional  scalar  curvature;
${\cal L}_m$ is  the  Lagrangian  of  the sources and  ${\cal G}$ denotes  the positive  determinant of  the  higher dimensional  space-time.  Since  gravitation  is  not  confined  to the four-dimensional  space-time,   the gravitational  coupling  constant  is not necessarily   the  same  as  that  of  General Relativity $k_g= 8\pi G$.   replaced  by   the  new coupling constant  $k_{g_*}$  to be determined from  high energy physics  experiments,  possibly  at  the Tev  scale  of  energies. For  consistency,     the   value   $k_g$  must be  restored  in  the  limit of  General Relativity.

The  functional variation  of \rf{EH}   with respect to the metric  ${\cal G}_{AB}$  gives the  D-dimensional Einstein's  equations
\be
  R_{AB} -\frac{1}{2}R {\cal G}_{AB}  =k_{g_*} T^*_{AB},\;\;  A,B=1..D  \label{eq:bulkEE}
\ee
where the    source term  in \rf{bulkEE}  represented  by  the  energy-momentum tensor  $T^*_{AB}$ is   composed by  known  observable sources in the    four-dimensional  space-time.

Equations  \rf{bulkEE}  are  again   of  the   hyperbolic  type, but  the  associated  topology  is  now   $I\!\!\R^4 \times  I\!\! \R^N$. Therefore,  just like  in the  ADM  3+1 metric  decomposition of  space-times,   the metric  geometry defined by \rf{bulkEE}  decomposes into  the  four-dimensional  space-time  components  and the  $N$ extra-dimensional  components.  To  find  these  components we  simply  write  the  metric solution of  \rf{bulkEE}  ${\cal G}_{AB}$   in  the    Gaussian  basis $\{\Z^A_{,\mu},\eta^A_a\}$ of  the embedding  space, obtained  from  equations  (\ref{eq:Z}-\ref{eq:Amuab}),  in the  form of  a  $(4+N)\times (4+N)$ matrix as:
\be
 {\cal G}_{AB}=
 \left(
 \begin{array}{cc}
            \tilde{g}_{\mu \nu}  + g^{ab}A_{\mu a} A_{\nu b}  & A_{\mu a}\\    A_{\nu b}  & g_{ab}
           \end{array}
           \right)  \label{eq:ansatz}
\ee
where we  have  denoted
\be
\tilde{g}_{\mu\nu}=\bar{g}_{\mu\nu}\!\!- 2y^{a}\bar{k}_{\mu\nu a}
+y^{a}y^{b}\bar{g}^{\alpha\beta}\bar{k}_{\mu\alpha
a}\bar{k}_{\nu\beta b}  \label{eq:tildeg}
\ee
These are   the  components  of  \rf{gmunu},   excluding the  terms involving   $A_{\mu ab}$, representing  the  metric  of  the  space-time   deformed   by  the extrinsic  curvature  alone.  The  components  $A_{\mu a}$ represent  just a  different  notation  for  cross  metric components $g_{\mu a}$ given by  \rf{gmua}.

As  it  is clear,  \rf{ansatz}  is   similar to the  metric   ansatz  used in the  standard non-Abelian Kaluza-Klein  theory. The   difference is  that  here  it  is a  direct  consequence of  the  $4+N$  hyperbolicity of  \rf{bulkEE}, which tells  that  at each point of  the  host space  we have  a four-dimensional  space-time and   $N$  orthogonal vector  fields.  Thus,  it  follows  from the  same  arguments that   the  Einstein-Hilbert  Lagrangian  \rf{EH}  written   for \rf{ansatz}  decomposes  into  the four dimensional Einstein-Hilbert  term  \rf{tildeg},  plus the Yang-Mills Lagrangian:
corresponding  to   the  third  fundamental form  $A_{\mu a b}$:
\be
\R\sqrt{\cal G}  =  \tilde{R}\sqrt{-{\tilde g}}  +  \frac{1}{4}tr {F}_{\mu\nu}{F}^{\mu\nu}\sqrt{-\tilde{g}}  \label{eq:KKLagrangian}
\ee
where   $F_{\mu\nu}= [D_\mu ,D_\nu]$,  and    $D_\mu=I\partial_\mu+  A_\mu$  are  defined  in the   Lie  algebra of  the (pseudo) rotations group  of the normal  vectors $\eta_a$.

Note  that  the   standard  basis  of  the Lie algebra of  a  rotation group  can be  expressed  with two indices
$L^{ab}$  such that   $[L^{ab}, L^{cd}]=  f^{abcd}_{mn} L^{mn}$,  where the factors  f  are  the  structure  constants.  In such basis  we  write
$A_\mu = A_{\mu a b}L^{ab} $  and
the  components  of  the  matrices
\be
D_{\mu a}{}^b  = \delta_a{}^b  \partial_\mu  +  A_{\mu a}{}^b \label{eq:D}
\ee
  On the  other hand, sometimes it is  more covenient to use the Killing basis of  the Lie  algebra with a  single  index  $K^a$,   to  write
$A_{\mu} = A_{\mu a}K^a$. The  final result is  independent  of  the  choice of  basis.

The    manifold  $\tilde{V}_4$  is   a  deformation of  $\bar{V}_4$,  produced   by    the  extrinsic  curvature  alone,  with  metric  is   \rf{tildeg},  defined  by  the   Lagrangian
\be
\tilde{\cal R}\sqrt{\tilde{g}}= [R\sqrt{g}-(K^2-h^2)]\sqrt{g} \label{eq:tildeL}
\ee
 In   general, the  final deformed  manifold  $V_{4}$ also  contains  the  contribution of  the third fundamental form. Its     geometric  components  are  all  given by  (\ref{eq:gmunu}-\ref{eq:Amuab})\footnote{In those  expressions  the  indices $ a, b, c \cdots$ above  are  raised  and  lowered  by  the  orthogonal metric  $g_{ab}$ and  the indices    $\mu,  \nu, \rho, \cdots$  are  raised  and  lowered  with the   metric of  the  $V_4$    foliation  and its  inverse.}.

Now  we may    prove   Ne'emann  conjecture: \textit{The  gauge  group  for the  field  $A_\mu$  defined  by third  fundamental form  is  the   group of  (pseudo)-rotations of  the    vectors  $\eta_a$ orthogonal  to the space-time}.\\
Indeed,  the  group of  rotations  of  the  orthogonal vectors     is    a  subgroup of  the group of  isometries of the embedding space  $\pounds_\xi \G_{AB}=0$.  Therefore,  consider an infinitesimal transformation of that subgroup,  given by
\[
x'^{\mu} = x^{\mu}, \hspace{1cm}  y'^{a}=y^{a} + {\xi}^{a}(x^{\mu},y^{a})
\]
 where the  descriptor  is  defined by  $\xi^{a}=\delta\theta_{b}^{a}(x^{\mu})y^{b}$ for  the infinitesimal parameters $\delta\theta_b^a(x^\mu)$,   functions of   the  space-time coordinates. The transformation of  each component  (A) of the  orthogonal vector field $\eta_a$ is given by
\[
\eta'^A_a(x)=\frac{\partial y^b}{\partial  y'^a}\eta^A_b (x)=(\delta^b_a-\xi^b_{,a})\eta^A_b
\]
Killing's  equation  for  the  entire  embedding space   $\xi_{(A,B)}=0$  gives  for the  considered  group $\xi_{(a,b)}=0$,  $\xi_{(\mu,\nu)}\equiv 0$,  with  solution $\xi^a = \delta\theta^a_m y^m$.

Applying the  above transformation to  the  definition of  $A_{\mu  ab}$ (given  by  \rf{kA}),   we  obtain
after  neglecting  second order  products of  $\delta \theta$:
\[
A'_{\mu ab} =\G_{AB}(\delta^m_a-\delta\theta^m_a)\eta^A_m [(\delta^n_b -\delta\theta^n_b)\eta^B_n]_{,\mu}  =A_{\mu ab}-2\delta\theta^n_{[b} A_{\mu a]n}  -\delta\theta_{ab,\mu}
\]
which is  the  typical  transformation of  a  gauge  potential  for  a  local gauge  group.

To complete  the  proof we  need  to show  that  the   third  fundamental form   satisfy   the  Yang-Mills  equations  for  the   considered   transformations.   For that purpose  we may use  the Lagrangian \rf{KKLagrangian},  or  more  appropriately,  the explicit  Lagrangian,  obtained by   separating  the    contributions  of   $g_{\mu\nu}$,  $k_{\mu\nu a} $  and  $A_{\mu a b}$  in   \rf{KKLagrangian}:
\be
{\cal L}  = [R +(K^2-h^2)]\sqrt{g}   -\frac{1}{4}tr F_{\mu\nu}F^{\mu\nu}\sqrt{g}  \label{eq:HDLagrangian}
\ee
where  the  determinant of  the  extra dimensional  metric is  a  common constant  factor  which   was  removed from  this Lagrangian.

Next   take  the    variation of the Lagrangian in \rf{HDLagrangian} with respect  to   $A_{\mu a b}$,
Since the \rf{HDLagrangian}   depends of  that  field  only in the term   $\frac{1}{4}trF_{\mu\nu}F^{\mu\nu}$,  the  variation of    it leads  to the   Yang-Mills equations
\begin{eqnarray}
&&  D_\mu  F^{\mu\nu}  =4\pi J^\nu \label{eq:ym1}   \\
&&  D_\mu  F^{*\mu\nu}=0  \label{eq:ym2}
\end{eqnarray}
where  $J^\mu$ is  the  Noether  current.
Noting that   the  components  $A_{\mu ab}$ are  written in  the  space  generated by the  rotations of  the vectors  $\eta_a$,  these  components  are  in the  adjoint representation  of   the   Lie  algebra  of  the  group of  rotations  of  the  vector  othogonal  to  $V_4$.

The  second  relevant result concerns  the   gravitational  equations.  These  can  also be  derived from  the  variations  of  \rf{HDLagrangian}  with respect  to the  three  components  $g_{\mu\nu}$, $g_{\mu a}$  and   $g_{ab}$.  Instead,   we  find  it  more illustrative  just to write  the higher-dimensional Einstein's  equations   \rf{bulkEE} in the Gaussian  basis  $\{ Z^A_{,\mu}, \eta^A_a \}$,  obtaining
\begin{eqnarray}
&&({\cal R}_{AB}-\frac{1}{2}{\cal R}{\cal G}_{ AB})\Z^A_{,\mu}\Z^B_{,\nu}=  R_{\mu\nu}-\frac{1}{2}R - Q_{\mu \nu} - T^{YM}_{\mu \nu} =\kappa_* T^*_{\mu \nu} \label{eq:G-tensor}\\
&&({\cal R}_{AB}-\frac{1}{2}{\cal R}{\cal G}_{ AB})\Z^A_{,\mu}\eta^B_{a}= k_{\mu a;\rho}^{\rho}\! -\!h_{a,\mu} \!+\! A_{\rho c a}k^{\rho \;c}_{\;\mu}\! -\!A_{\mu c a}h^{c}
 \!  = \kappa_*   T^*_{\mu a}   \label{eq:G-vector}\\
&&({\cal R}_{AB}-\frac{1}{2}{\cal R}{\cal G}_{ AB})\eta^A_{a}\eta^B_{b}= \frac{1}{2}[R-(K^{2} -h^{2})]g_{ab} =\kappa_* T^*_{ab}  \label{eq:G-scalar}
\end{eqnarray}
where  we  have  denoted
\be
Q_{\mu\nu}  = g^{ab}(k^{\rho}_{\mu a} k_{\rho \nu b} -h_a k_{\mu\nu b}) -\frac{1}{2}(K^2-h^2)g_{\mu\nu} \label{eq:Qmunu}
\ee
Here  $h_{a}= g^{\mu\nu}k_{\mu\nu a}$  represents  the   mean  curvature of the four-dimensional space-time with respect to the  $\eta_a$  direction.  If  we  consider all  extra  dimensions,   we  obtain  $h^2=g^{ab}h_a h_b$.  The  Gaussian  curvature of  the space-time  is  $K^2= k^{\mu\nu a}k_{\mu\nu a}$.
The   last term in the left  hand  side term  in  \rf{G-tensor} is  due  to the  presence  of  $A_{\mu ab}$  in  \rf{KKLagrangian}, corresponding to    the   energy-momentum  tensor of the gauge (Yang-Mills)  field  built  from  the components     $A_{\mu ab}$:
\[
T^{YM}_{\mu \nu} = (F_{\mu}^{ \alpha}F_{\nu}^{ \beta} g_{\alpha \beta} - \frac{1}{2}g_{\mu \nu} F^{\alpha \beta}F_{\alpha \beta})
\]
 Finally,  $T^*_{\mu\nu}$,  $T^*_{\mu a}$,  and  $T^*_{ab}$  are the projections of  $T^*_{AB}$  on the  tangent, cross   and  the normal  directions of  the  space-time.  Obviously,  admitting  that   these  are  composed of  ordinary matter   they are  confined  and  conserved  in the  four-dimensional space-time,  so  that  it is  natural  to  assume  that  $T^*_{\mu a}=0$ and  $T^*_{ab}=0$.  In this  case,   equations   \rf{G-vector} and  \rf{G-scalar} become  homogeneous  and  consequently   the  extrinsic  curvature  cannot be  completelly  determined,  requiring an  additional  equation.

\section{The  Spin-2 Extrinsic  Curvature}

On the physical  side,  the  extrinsic  curvature corresponds  to a  spin-2  field  for  each internal index, or  more appropriately,  to  a  multiplet  of  spin-2 fields,  which  are  independent of  the metric  in each  four-dimensional  space-time  of  the  foliation. Several  massive   spin 2 particles are  known,    composing    a  nonet  as  for  example  the   f, $A_2$  and the $K^*$  mesons.  The  gravitational  field  is  also a  spin-2  field  but  since  it  has  a long range,  it must be    a massless field.
A Known  theorem    due  to    Soraj Gupta   in  1960  and later reviewed   by Stanley Deser an others in 1970,  proved  that  \textit{ any  massless   spin-2  field  defined by a  symmetric  rank-2  tensor is necessarily a  solution of  an  Einstein-like  equations}.
  The  Gupta theorem   essentially  reverses  the      linear approximation of  Einstein  equations,   applied to the  masslless  Fierz-Pauli  theory,  in practice  reconstructing    the non-linear terms  \cite{Gupta,Deser}.

 In  1971,   C. Isham   and  others,  proposed   that  one such massive  spin-2  field    would  act   as  an intermediate  field  between   Einstein's  gravity and  hadrons,  as  a solution of  Gupta's  equation \cite{Salam}. During that  same  period,  the  existence  of  a    short range  gravitational   field  with mass was  considered  as  a possible   modification of  General Relativity, starting  from the  a  non-linear  equation derived from  the   Pauli-Fierz spin-2  action  with mass,  so that    General relativity  would  be  recovered  in the  zero mass limit
\cite{vDam,Zakharov}.  However,  it was soon   found  that in this  limit the  theory   gives  a  different theory  containing ghosts  and   that does not agree  with the observed  gravitational light  bending  experiment \cite{Boulware},  although  it was argued  that  this  could be    corrected  if  the  non-linear terms  in the Fierz-Pauli equation   would  be  taken into account  in a  special parametrization \cite{Vainshtein}.  Such possibility  led  to a  renewed  interest  in the  construction of    massive gravity  in a  four-dimensional  theory  with two  independent metrics (a bi-metric theory), each  one  responding  to  a  different   set   of  field  equations.  However, there  are  stil  some  issues  as it  has been  argued  that  one   of  these  equations  would   constrain the  other  \cite{Visser}.

In the  following  we  use  a   different  approach,  where    the  extrinsic  curvature  of  the space-time   plays  the role  of  such   massive  gravitational  field.  As we  have  seen,  the  existence  of  the  extrinsic   curvature   is  essential  for  the  inclusion of gravitation in the  standard  model of  unification,  as  well as  to  the  explanation of  the  acceleration of  the universe \cite{MaiaGDE}.

The  Fierz-Pauli  Lagrangian  for  a   relativistic spin-2 field described  by a  symmetric  rank  two  tensor  field $H_{\mu\nu}$ in Minkowski's  space-time\footnote{ The  Lagrangian  in a  curved  space-time with a  non-minimal coupling with Einstein's  gravitation has been considered, but it is  dependent on what  type of additional  terms  should be included\cite{Gitman}.} is
\begin{equation}
{\cal L} =\frac{1}{4}\left[ H_{,\mu}H^{,\mu} -H_{\nu\rho,\mu}H^{\nu\rho,\mu}  -2H_{\mu\nu}{}^{,\mu}H^{,\nu}+ 2 H_{\nu\rho,\mu}H^{\nu\mu,\rho}- m^2(H_{\mu\nu}H^{\mu\nu} -H^2)\right]  \label{eq:Fierz}
\end{equation}
where  $H=\eta^{\mu\nu}H_{\mu\nu}$  and  $m$ is  a constant. Assuming   that   $H_{\mu\nu}$  is     trace free: $H=0$, and  that it is  divergent free:  $H^{\mu\nu}{}_{,\nu}=0$,   we  obtain the  Klein-Gordon  equation  $(\Box^2 -m^2)H_{\mu\nu}=0$ so that   $m$ can be  interpreted  as   the Klein-Gordon  mass  of   the spin-2 field $H_{\mu\nu}$ \cite{FierzPauli}.

Applying the above  Lagrangian  to the  extrinsic  curvature  $k_{\mu\nu a}$  in the  curved  space-time   $V_4$  with  metric  $g_{\mu\nu}$,  with  minimal  coupling  with gravitation, the  Fierz-Pauli Lagrangian becomes
\begin{eqnarray}
{\cal L}&&=\frac{1}{4}  \left[ h_{a, \mu}h^{a,\mu}-k_{\rho\nu a ;\mu} k^{\rho\nu a; \mu}- 2k_{\mu\nu a}{}^{;\mu} h_a{}^{,\nu}  +2 k_{\nu\rho a ;\mu} h^{a;\rho}
 -m^2(K^2-h^2)  \right]
\end{eqnarray}
Where  the  semicolon  denotes  the covariant derivative  with respect to the  space-time  metric  $g_{\mu\nu}$.
The  Euler-Lagrange  equations  with respect to  $k_{\mu\nu a}$ are
\begin{eqnarray}
&&\Box^2 k_{\mu\nu a} -g_{\mu\nu}\Box^2 h_a +h_{a , \nu ;\mu}  +g_{\mu\nu}k_{\alpha\beta a}{}^{;\alpha\beta}- k_{\sigma\mu a ;\nu}{}^{;\sigma}
-m^2(k_{\mu\nu a} - h_a g_{\mu\nu})=0
\end{eqnarray}
Here  $\Box^2$  satands  for   the   covariant  D'alambertian operator.

The Klein-Gordon mass   of  the  extrinsic  curvature appear  under two
conditions   $h_a=  g^{\mu\nu}k_{\mu\nu a}=0$  and  $k_{\mu\nu a}{}^{;\nu}=0$:
\be
(\Box^2-m^2)k_{\mu\nu a}=0
\ee
 Since   the  space-time  coordinates  and  the  extra  dimensional  coordinates  are independent,  the second condition  $k_{\mu\nu a}{}^{; \nu}=0$  becomes  trivial  in view  of  \rf{York}.  On the other hand,  $h_a$ =0 is  more  specific,  telling that   the  space-time  is  minimal  in the   sense of  minimal area surfaces.
 In the case of an  embedded Riemannian manifold,  this   occurs  when  the extrinsic  curvature is   totally intrinsic,   proportional  to the metric:
$k_{\mu\nu a}= \alpha_a g_{\mu\nu}$.  However  from  \rf{G-tensor}  it follows  that   $Q_{\mu\nu}= \Lambda g_{\mu\nu}$  where  $\Lambda$ is proportional  to   the cosmological  constant and  also proportional to the inverse of  the  curvature  radius  of  the  space-time.   Therefore, we  conclude  that   $h_a=0$  corresponds  to  the  infinite limit of  the curvature  radius  which   means   that  the Klein-Gordon mass  term is  characterized  in the  Minkowski space-time  by  the  mass   operator of  the  Poincar\'e  group  as  one  would  expect.
\vspace{3mm}\\
\textbf{Summary}:\vspace{3mm}\\
The  current effort  to   define a  short range  component of  the  gravitational  field  has  been  motivated  by    the  necessity to  understand  the  acceleration of  the universe  as  a  correction to  Einstein's  gravity,  but also by   high energy  physics,  notably to  obtain  an  intermediator  between  Einstein's massless   gravitation  and  the  gauge and  matter  fields.  However,  as it  was pointed out   Einstein's  gravity  in four  dimensions  seems  to be  a  unique  theory, that is  a  theory by  its  own, and  not a limit  of  some   massive   spin-2  theory  \cite{Deser-Damour,Visser}.

 On the other hand,   considerations  in   higher  dimensional  gravity  has  shown  that it is possible  to break  the gravitational  hierarchy,  at the same  time  that gravitation  acquire  greater degrees  of  freedom.  In  particular,   Kaluza-Klein  theory   splits  the  higher dimensional metric  in  four-dimensional gravitation  plus   four-dimensional  Yang-Mills   gauge  fields.   A  nice    theory  that   did not  work  because  of  an  ill  justified  geometrical  construction  based on the   product  topology   $I\!\!R^4  \times  B_n$,  where  $B_n$ is  a  small compact  space.

Based  on  a  proposition of   Joseph and  Ne'emann   from the early 60's  and  on the Nash  theorem  from  mid  60's  and  on the  discussion of   massive  gravity  from  the early  70's  to  the present day,  we  have  completely  revised  Kaluza-Klein  theory,  by removing the  compact  space,  introducing the rather   successful  theory of  smooth  manifold  deformations,  and   using   the  extrinsic  curvature  as  a  massive  spin-2  field  that  couples  with  Einstein's  gravitation and  with  gauge  fields,  and  generalize  the   cosmological constant  term  $\Lambda g_{\mu\nu}$   to  a  cosmic  tensor  $ Q_{\mu\nu}$  built  with the extrinsic  curvature. The  cosmological  constant case  corresponds  to  the  intrinsic  limit  of  the theory,  where  the  extrinsic  curvature becomes  intrinsic, proportional to the metric.  It is  also  interesting  to notice  that   since  the  extrinsic  curvature   referes  to  the propagation of  the metric  in the  extra dimensions,   the  eventual ghost  states   will be  outside the space-time.

 The   gauge  group must be  detailed  by  a    phenomenological   analysis  of  the  symmetries  required  by  the  GUT  scheme \cite{Slansky}. Since      Nash's  embedding  theorem  for  space-times  can  be  implemented    with     $D= 14$  dimensions,  this  result  may point  to   GUT  based on a  45  parameter  group   like  for  example   $SO(10)$ or   equivalent.  It is  also  possible  to  look  for  a  larger  gauge symmetry  containing  those  45 parameter  groups.

It  is  interesting  to  note  that  the above proposed  unification  occurs  only  when  we have  more  than  one  extra  dimension   (That  is,  six  dimensions altogether). This  is  a   limiting  case where  the  gauge  group  is  either  $SO(2)$ or  $SO(1,1)$. The   first case corresponds  to  the unification  of   gravitation  with  the  electromagnetic  field.  This  happens  for  example  when  we  consider  the  Kruskal  space-time and  the second case  corresponds  to the Schwarzschild  space-time  (up to the  horizon).  In both cases  the  gravitational  field are   static  so  that  the  gauge  fields  $A_{\mu ab}$ and  the strong  gravitational  field  $f_{\mu\nu a}$  are also static.
Therefore,   in the   $ SO(2) \sim SU(2)$  case  we  obtain a  very simple unification  of the Schwarzschild   gravitation  with  the  electrostatic  field,  corresponding  to  the  existence  of  a   static  charge  located at the  point  $r=o$  which  is where the  singularity of  the  metric  is  located.  Even  in the  vacuum  case,  the  two above cases do not follow from  not solutions  of  the  Einstein-Maxwell equations in  four-dimensions  due  to the presence of  the   tensor  $Q_{\mu\nu}$ representing the  conserved  energy  of  the  massive   gravitational  field at  Tev  energy  scale. A  more interesting example  of  the unification  may  be constructed  with  the  embedding of  the Kerr  space-time where  magnetic  field  would  also appear.

\vspace{1cm} ~\\
\textbf{Acknowledgements}:

The  authors  would like  to express  their  thanks  to Dr. Renato Portugal  for  his  help  in setting a  Maple  program  for  the  geometry of  embedded  Riemannian  submanifolds.

\end{document}